\documentclass[aps,prd,showpacs,twocolumn]{revtex4-1}

\usepackage{bm}
\usepackage{amsmath}
\usepackage{graphicx}
\usepackage{subfigure}
\usepackage{color}
\usepackage{mathtools}

\renewcommand{\vec}{\bm}

\def\ZB{Zitterbewegung}
\def\ZB{ZB}
\def\sumint{\mathclap{\displaystyle\int_p}\mathclap{\textstyle\sum}}

\begin{document}

\title{Zitterbewegung of neutral relativistic particles in static longitudinal fields}

\author{Tihomir G. Tenev}
\affiliation{Department of Physics, Sofia University, 5 James Bourchier Blvd, Sofia 1164, Bulgaria}
\author{Nikolay V. Vitanov}%
\affiliation{Department of Physics, Sofia University, 5 James Bourchier Blvd, Sofia 1164, Bulgaria}%

\date{\today}

\begin{abstract}
Zitterbewegung of neutral relativistic particles propagating along a constant magnetic and/or electric field is studied.
It is shown that spin Zitterbewegung, when superimposed on the Larmor precession frequency, leads to a beating pattern.
The existence of a forbidden frequency of spin precession is predicted.
Modifications of position and velocity Zitterbewegung due to lifted spin degeneracy manifested in the appearance of longitudinal and transversal Zitterbewegung, each with two Zitterbewegung frequencies and resulting beating patterns, are reported.
\end{abstract}

\pacs{03.30.+p, 03.65.Pm, 31.30.J-, 03.67.Ac}
\keywords{Zitterbewegung, Dirac equation, special relativity, Larmor precession, spin splitting}

\maketitle

\section{Introduction}
The Dirac equation unified special relativity with quantum mechanics for fermions.
It was a step further beyond the Schr\"odinger equation due to its invariance with respect to Lorentz boosts.
Several early experimental successes of it were the accurate prediction of the intrinsic magnetic moment of the electron,
 the prediction of antiparticles and the correct description for the fine-structure of the spectrum of the hydrogen atom.
These were based on a form of the Dirac equation which describes relativistic electrons.
There exists a similar form of it, which offers a phenomenological description of some compound particles like the neutron and the proton in the relativistic domain.

Despite its successes regarding the relativistic behavior of the electron, certain solutions of the Dirac equation predict some unusual effects, which have not been experimentally measured so far for neither the electron, the neutron or any other particle~\cite{Thaller}.
The best known of them are (i) Klein's paradox~\cite{Thaller} --- the tunneling of a relativistic electron into an infinite potential barrier, and (ii) Zitterbewegung (ZB) \cite{Thaller} --- helicoidal motion of the expectation value of the position $\hat{\vec{r}}$ and oscillatory behavior of the expectation values of the momentum $\hat{\vec{p}}$, the spin $\hat{\vec{S}}$ and the orbital angular momentum of a free electron.
Recently these two effects have gathered renewed attention by the scientific community, thanks to several proposals for their experimental emulation in physical systems like solid-state quantum wells~\cite{Schliemann05}, trapped ions~\cite{Lamata,Casanova,Bermudez,Gerritsma}, graphene~\cite{ZawadskiGraphene} and optical waveguides~\cite{Longhi}.

Most of the recent work on \ZB~is focused on electrons.
In contrast here we consider the Dirac equation for neutral particles propagating relativistically in external magnetic and/or electric fields~\cite{Thaller} along the direction of the applied field.
The Dirac equation~\cite{Thaller} for these particles deviates from the Dirac equation for the electron in three aspects: (i) the term coupling the electrostatic potential to the charge of the particle dissapears; (ii) the terms coupling the vector potential to the charge and momentum are not present;  (iii) a term coupling the anomalous magnetic dipole moment (MDM) to the magnetic field and a term coupling the eventual electric dipole moment (EDM) to the electrostatic field appear.

\section{\ZB~of free particles}

The Dirac Hamiltonian for a neutral particle in the absence of external fields reads
\begin{equation}
\hat{H} = c\hat{\alpha}_x\!\cdot\!\hat{p}_x + \hat{\beta} m_0c^2\ ,\label{EQ:FreeHam}
\end{equation}
where $c$ is the speed of light, $m_0$ is the rest mass of the particle, $\hat{\alpha}_x$ and $\hat{\beta}$ are the Dirac matrices.
The spectrum of the Hamiltonian is doubly degenerate and its eigenvalues are $E_{\pm}=\pm E$, with $E=\sqrt{(cp_x)^2+(m_0c^2)^2}$.
It is well known~\cite{Thaller,AdvancedSakurai} that if the initial state is a superposition of both positive and negative energy eigenstates, in the process of evolution the expectation values of velocity,  position and spin will oscillate with the same frequency
\begin{equation}\label{free-ZB}
\omega^{\text{zb}}=\frac{2E}{\hbar}=\frac{2}{\hbar}\sqrt{(cp_x)^2+(m_0c^2)^2}.
\end{equation}
This is the free-particle \ZB.

\begin{figure}
\begin{center}
\includegraphics[scale=0.7]{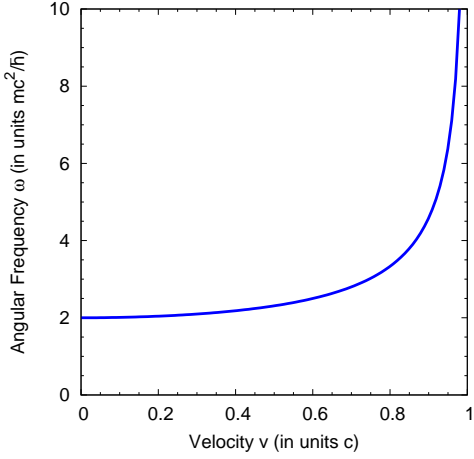}
\end{center}
\caption{Demonstration of blue shift effect of free particle \ZB~frequency $\omega^{\text{zb}}$, Eq.~\eqref{free-ZB}. 
The \ZB~frequency $\omega^{\text{zb}}$ increases with the particle velocity $v$, and respectively, its momentum $p=\gamma mv$, $\gamma=1/\sqrt{1-(v/c)^2}$.}\label{FIG:WZB}
\end{figure}

There exists a lower bound  $\omega=2mc^2/\hbar$ for the free-particle \ZB~of the three quantities $\hat{\vec{S}}$, $\hat{\vec{\alpha}}$ and $\hat{\vec{r}}$.
In the literature~\cite{Zawadski} it is usually accepted that the frequency of the free-particle \ZB~is approximately $\omega=2mc^2/\hbar$.
However, for relativistic speed of propagation, $cp_x$ becomes comparable to $mc^2$.
Then there will be considerable contribution of $cp_x$ to the \ZB~frequency $\omega^{\text{zb}}$, Eq.~\eqref{free-ZB}. 
The difference
\begin{equation}
\Delta\omega^{\text{zb}}=\frac{2}{\hbar}\sqrt{(cp_x)^2+(m_0c^2)^2}-\frac{2mc^2}{\hbar}\; ,
\end{equation}
between the rest frame ($p_x=0$) and the observer frame of reference amounts to a \emph{blue shift} of the \ZB~frequency: it is demonstrated in Fig.~\ref{FIG:WZB}.
Therefore, the free-particle \ZB~frequency has the opposite behavior of the earlier reported~\cite{TenevB} relativistic red shift effect for the frequency of Larmor precession.
It is a consequence of the relativistic energy-momentum-mass relationship $E^2=(pc)^2 + (mc^2)^2$, and the fact that \ZB~is observed due to an interference between positive and negative energy solutions.

\section{Modification of \ZB~due to lifted spin degeneracy}

Both Larmor precession of spin along an external field and spin-\ZB~manifest themselves as spin precession.
However, despite their similarity there are important differencies between them.
Spin-\ZB~is a consequence of interference between positive and negative energy components of the Dirac equation, while Larmor precession is a consequence of lifting of spin-degeneracy by external fields.
It is natural then to expect some modifications of at least spin-\ZB, and eventually position-\ZB, when spin degeneracy is lifted by external fields in the Dirac equation.

We address these problems by examining a relativistic model of a neutral particle possessing MDM, and eventually EDM, placed in an external field, in which the static magnetic and/or electric fields are applied along the propagation direction.
Working with the $x$-direction in SI units the Hamiltonian takes the form
\begin{equation}
\hat{H} = c\hat{\alpha}_x \hat{p}_x + \hat{\beta}mc^2 + 2\hat{\beta}\hat{S}_x\left(d E - \mu B\right)  \; , \label{EQ:1DRelHam}
\end{equation}
where $\hat{S}_x$ is the $x$-component of the spin vector operator in relativistic theory, $E$ is the electric field, $B$ is the magnetic field, $d$ is the size of the EDM and $\mu$ is the size of the MDM of the relativistic particle.
The four distinct eigenvalues of the Hamiltonian are
\begin{subequations}\label{EQ:EigenE}
\begin{eqnarray}
E_{\pm}^{\uparrow}   = \pm \sqrt{(cp)^2  + (mc^2 + \Delta)^2}\; ,  \\
E_{\pm}^{\downarrow} = \pm \sqrt{(cp)^2  + (mc^2 - \Delta)^2}\; ,
\end{eqnarray}
\end{subequations}
where \mbox{$\Delta = d E -\mu B$}. The four distinct eigenvalues of the Hamiltonian correspond to four different frequencies in the matrix elements between the eigenstates,
\begin{subequations}
\begin{eqnarray}
 \omega_{\text{L}}&=& \frac{1}{\hbar}(E_{+}^{\uparrow}-E_{+}^{\downarrow})= \frac{1}{\hbar}(E_{-}^{\downarrow}-E_{-}^{\uparrow})\; , \\
 \omega^{\text{zb}}_1 &=&  \frac{1}{\hbar} (E_{+}^{\uparrow}-E_{-}^{\uparrow})\; ,  \\
 \omega^{\text{zb}}_2 &=&  \frac{1}{\hbar} (E_{+}^{\uparrow}-E_{-}^{\downarrow}) =  \frac{1}{\hbar} (E_{+}^{\downarrow}-E_{-}^{\uparrow})\; , \\ 
 \omega^{\text{zb}}_3 &=&  \frac{1}{\hbar} (E_{+}^{\downarrow}-E_{-}^{\downarrow})\; .
\end{eqnarray}
\end{subequations}
Using the plane wave eigenstates $|l,s\rangle$  of the Hamilatonian (\ref{EQ:1DRelHam}), an arbitrary wavepacket can be represented as the superposition
\begin{equation}\label{EQ:WavePacket}
|\Psi\rangle = \;\;\sumint\;\; \sum_{s=\uparrow,\downarrow}\sum_{l=\pm}c_{l,p,s}|l,s\rangle e^{\frac{\text{i}}{\hbar}(p\cdot x-E_{l}^{s})}, 
\end{equation}
which includes both positive and negative energy eigenvectors~\cite{AdvancedSakurai}.

\subsection{Modification to spin-\ZB}

The expectation value $\langle\Psi|\hat{S}_x|\Psi\rangle$ of the component of the spin vector operator along the propagation direction does not contain an oscillating component.
Its value $\langle\Psi|\hat{S}_x|\Psi\rangle=\;\sumint\;\left( \sum_{l=\pm}\sum_{s=\uparrow,\downarrow}|c_{p,l,s}|^2\langle s,l|\hat{S}x|s,l\rangle \right)$ is a constant of motion, determined by the initial conditions.
This is natural since in this geometry $\hat{S}_x$ is the helicity of the particle.
The expectation value of the $y$ and $z$ components of the spin operator read ($j=y,z$)
\begin{align}
\langle\hat{S}_j\rangle & = \;\;\sumint\;\; \sum_{l=\pm} 2\text{Re}(c^*_{p,l,\uparrow}c_{p,l,\downarrow}\langle\uparrow,l| \hat{S}_j|l,\downarrow\rangle e^{l i\omega_{L}t}) + \nonumber \\
&+ \;\;\sumint\;\; \sum_{s\neq s'} 2\text{Re} ( c^*_{p,+,s'}c_{p,-,s}\langle s',+|\hat{S}_j|-,s\rangle e^{i\omega_{\text{zb}}^2t} ).
\end{align}
They contain two oscillating terms with different frequencies: a relativistic Larmor precession frequency,
\begin{align}
\omega_L &=\frac{1}{\hbar}\sqrt{(cp)^2  + (mc^2 + \Delta)^2}\nonumber\\
&-\frac{1}{\hbar}\sqrt{(cp)^2  + (mc^2 - \Delta)^2},
\end{align}
 and a modified spin-\ZB~frequency,
\begin{align}
 \omega_2^{\text{zb}} &= \frac{1}{\hbar}\sqrt{(cp)^2  + (mc^2 + \Delta)^2} \nonumber\\
  &+\frac{1}{\hbar}\sqrt{(cp)^2  + (mc^2 - \Delta)^2}.
\end{align}
The presence of two different frequencies for every plane wave component leads to a beating pattern with an angular frequency
\begin{equation}
\omega^{\text{sb}}=\omega_2^{\text{zb}}-\omega_L=\frac{2}{\hbar}\sqrt{(cp)^2+(mc^2-\Delta)^2}.
\end{equation}

Earlier~\cite{TenevB} we have reported the existence of an upper bound $2mc^2/\hbar$ for the Larmor precession frequency of a relativistic neutral particle.
On the other hand, the lower bound for the free-particle spin-Zitterbwegung is not modified by the coupling of EDM and MDM to external fields and therefore it is still $2mc^2/\hbar$.
However, in a realistic situation the two limits can not be achieved by $\omega_L$ and $\omega_2^{\text{zb}}$ due to the Heisenberg uncertainty principle. Thus the frequency
\begin{equation}
\omega_{\text{forbidden}}=\frac{2mc^2}{\hbar}
\end{equation}
is a forbidden frequency for spin-precession within the considered model.
Below it, spin precession is due to Larmor precession, whereas above it, it is because of \ZB.
For the Hamiltonian~\eqref{EQ:1DRelHam} the transverse spin components precess with the two frequencies at the same time leading to the beating pattern with beating frequency $\omega^{\text{sb}}$.

\begin{figure}
\begin{center}
\includegraphics[scale=0.7]{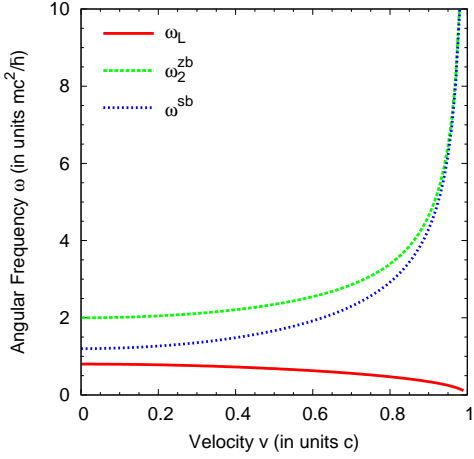}
\end{center}
\caption{Demonstration of blue shift effect for the spin-\ZB~frequency $\omega^{\text{zb}}_2$, red shift for Larmor precession frequency $\omega_L$ and blue shift effect for their difference - the spin beating frequency $\omega^{\text{sb}}$. The figure is plotted assuming $\Delta=0.4\; mc^2$}\label{FIG:SB}
\end{figure}

The modified \ZB~frequency $\omega_2^{\text{zb}}$ exhibits a relativistic blue-shift effect similar to the one discussed for free-\ZB~frequency and contrary to the Larmor precession frequency\cite{TenevB}, which is subjected to red-shift effect.
The beating frequency  $\omega^{\text{sb}}$ also exhibits blue shift effect --- it is larger for particles with larger velocity and respectively momentum.
The behavior of the three frequencies versus the velocity is illustrated in Fig.~\ref{FIG:SB}.

\subsection{Modifications to velocity and position \ZB}

While modification of spin-Zitterbwegung by MDM-B or EDM-E coupling is expected, the modification of orbital dynamics by the coupling of an external magnetostatic/electrostatic field to the spin-degree of freedom is more surprising.
We investigate these by computing the expectation values of $\hat{\alpha}_j$ and $\hat{r}_j$ ($j=x,y,z$) with respect to the superposition~\eqref{EQ:WavePacket}.
We first calculate the modified velocity-\ZB~in the Schr\"odinger representation and then the position-\ZB~from the expression
\begin{equation}
\frac{d}{dt}\int\Psi^+(x,t)\hat{x}\Psi(x,t)d^3x = c\int\Psi^+(x,t)\hat{\alpha}_x\Psi(x,t)d^3x
\end{equation}
 and integration~\cite{AdvancedSakurai} of the expectation value of velocity with respect to time and multiplication by $c$.

It turns out that Eq.~\eqref{EQ:1DRelHam} leads to two different types of velocity and position \ZB:
 (a) \emph{longitudinal} for the $x$ component of the velocity and position operators here,
 and (b) \emph{transverse} for the $y$- and $z$- components of the velocity and position operators.
They are characterized by different precession frequencies.

The expectation value $\langle\alpha_x\rangle$ for the longitudinal velocity-\ZB~reads
\begin{eqnarray}\label{EQ:VZSDL}
\langle\hat{\alpha}_x\rangle &=&\;\;\sumint\;\; \sum_{l,s} \frac{cp_x}{E^s_{p,l}} +
 \;\sumint\;\; N_1(p_x) 2\text{Re}\left[c^*_{+,p,\uparrow}c_{-,p,\uparrow}e^{i\omega_1^{\text{zb}}t}\right] + \nonumber \\
 &+&\;\;\sumint\;\; N_2(p_x) 2\text{Re}\left[c^*_{+,p,\downarrow}c_{-,p,\downarrow}e^{i\omega_3^{\text{zb}}t}\right]\; ,
\end{eqnarray}
where
$N_1(p_x)=\frac{E_0^{\uparrow}cp_x}{\sqrt{E_+^{\uparrow}E_-^{\uparrow}(E_+^{\uparrow}+E_0^{\uparrow})(E_-^{\uparrow}+E_{0}^{\uparrow})}}$, $E_0^{\uparrow}=mc^2+\Delta$ and $N_2(p_x)=\frac{E_0^{\downarrow}cp_x}{\sqrt{E_+^{\downarrow}E_-^{\downarrow}(E_+^{\downarrow}+E_0^{\downarrow})(E_-^{\downarrow}+E_{0}^{\downarrow})}}$,
$E_0^{\downarrow}=mc^2-\Delta$.
The first term of Eq.~\eqref{EQ:VZSDL} describes the group velocity of the wavepacket, obtained as the weighted sum of the averaged velocities of the components of the wavepacket~\eqref{EQ:WavePacket}.
The other two terms describe the velocity-\ZB~for the Hamiltonian~\eqref{EQ:1DRelHam}.
Excluding the initial position, the expectation value of the transverse velocity-\ZB~reads
\begin{eqnarray}\label{EQ:PZSDL}
\langle\hat{r}_x\rangle &=&\;\;\sumint\;\;\sum_{l,s} \frac{cp_x}{E^s_{p,l}}c t +
 \;\;\sumint\;\; N_1(p_x) 2\text{Re}\left[\frac{c^*_{+,p,\uparrow}c_{-,p,\uparrow}}{i\omega_1^{\text{zb}}}e^{i\omega_1^{\text{zb}}t}\right] + \nonumber \\
 &+& \;\;\sumint\;\; N_2(p_x) 2\text{Re}\left[\frac{c^*_{+,p,\downarrow}c_{-,p,\downarrow}}{i\omega_3^{\text{zb}}}e^{i\omega_3^{\text{zb}}t}\right]\; ,
\end{eqnarray}
where again the first term describes classical motion while the other two describe position-\ZB.

\begin{figure}
\begin{center}
\includegraphics[scale=0.7]{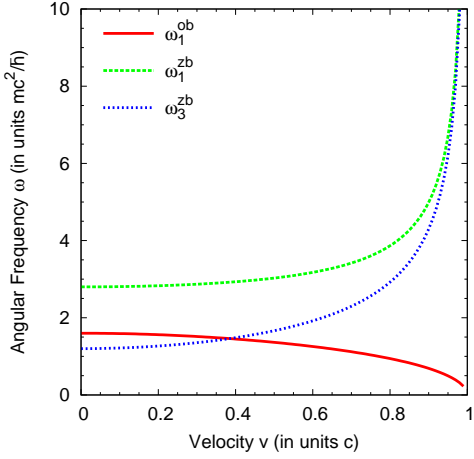}
\end{center}
\caption{Demonstration of blue shift effects fir  longitudinal velocity and position \ZB~frequencies $\omega^{\text{zb}}_{1}$ and $\omega^{\text{zb}}_3$, and red shift effect for their difference - the orbital beating frequency $\omega^{\text{ob}}_1$. Plotted assuming $\Delta=0.4\; mc^2$. Clearly the longitudinal \ZB~frequency $\omega^{\text{zb}}_1$ can take values below $2mc^2/\hbar$, as well as the \ZB~beating frequency $\omega^{\text{ob}}_1$.}\label{FIG:LZ}
\end{figure}

In contrast to the free particle case~\cite{Thaller,AdvancedSakurai} where there is just one frequency for velocity- and position-\ZB, in Eq.~(\ref{EQ:VZSDL}) and Eq.~(\ref{EQ:PZSDL}) there are two different frequencies of longitudinal velocity-\ZB~$\omega^{\text{zb}}_1=2\sqrt{(cp)^2+ (mc^2+\Delta)^2}$ and $\omega^{\text{zb}}_3=2\sqrt{(cp)^2+(mc^2-\Delta)^2}$. Both of them are different from the free particle \ZB~frequency $\omega^{\text{zb}}=2\sqrt{(cp)^2+ (mc^2)^2}$, by a term determined by the size of the spin-splitting $\Delta$.
In the rest frame of the particle they become
\begin{subequations}
\begin{eqnarray}\label{EQ:LongiFreq}
\omega^{\text{zb}}_1(0)=\frac{2mc^2}{\hbar} + \frac{2\Delta}{\hbar}\; , \\
\omega^{\text{zb}}_3(0)=\frac{2mc^2}{\hbar} - \frac{2\Delta}{\hbar}\; . \label{EQ:LongiFreqB}
\end{eqnarray}
\end{subequations}
Obviously,  $\omega^{\text{zb}}_1$ is always greater than the lower limiting frequency of free-particle \ZB~${2mc^2}/{\hbar}$, while $\omega^{\text{zb}}_2$ can take values which are smaller;
 this is illustrated in Fig.~\ref{FIG:LZ}.
In principle, by application of strong enough magnetic field the longitudinal velocity/position-\ZB~frequency can be lowered to the observable range for current technology.
Thus when the spin degeneracy of the particle is lifted by an external field,  ${2mc^2}/{\hbar}$ is not a lower limit for velocity/position-\ZB, while it is still a lower limit for spin-\ZB, as shown above.

Furthermore, the presence of two different frequencies in the expressions for longitudinal velocity/position-\ZB~leads to the appearance of a beating pattern with an orbital beating frequency
\begin{align}
\omega^{\text{ob}}_1 = \omega^{\text{zb}}_1 - \omega^{\text{zb}}_3 &= 2\sqrt{(cp)^2+ (mc^2+\Delta)^2} \nonumber\\
 &- 2\sqrt{(cp)^2+(mc^2-\Delta)^2} \; .
\end{align}
The beating frequency is just twice the Larmor precession frequency of spin-splitting $\omega^{\text{ob}}_1=2\omega_{\text{L}}$, which makes the predicted effect readily observable with current technology.
With respect to Lorentz transformations of the coordinate systems the two longitudinal velocity/position \ZB~frequencies exhibit \emph{blue shift effect} similar to the one for the spin-\ZB~frequency.
In contrast, the beating frequency, as their difference, manifests relativistic \emph{red shift effect}.
This is the opposite behavior of the beating frequency $\omega^{\text{sb}}$ of spin-\ZB.
These dependences of the \ZB~frequencies on velocity are plotted on Fig.~\ref{FIG:LZ}.

The expectation values of the $y$ and $z$ components of velocity/position have similar structures as the expectation values of spin-\ZB.
They have terms with two frequencies $\omega_L$ and $\omega^{\text{zb}}_2$, which are different from the free-particle \ZB~and from the longitudinal velocity/position-\ZB~frequencies $\omega^{\text{zb}}_1$ and $\omega^{\text{zb}}_3$.
The position expectation values reads
\begin{align}
\langle\hat{r}_j\rangle & =\;\;\sumint\ \ \sum_{l=\pm} 2\text{Re}\left[\frac{c^*_{p,l,\uparrow}c_{p,l,\downarrow}}{l\text{i}\omega_L}\langle\uparrow,l| \hat{\alpha}_j|l,\downarrow\rangle e^{l i\omega_{L}t} \right] + \nonumber\\
&+\;\;\sumint\ \ \sum_{s\neq s'} 2\text{Re}\left[\frac{c^*_{p,+,s'}c_{p,-,s}}{\text{i}\omega_2^{\text{zb}}}\langle s',+|\hat{\alpha}_j|-,s\rangle e^{i\omega^{\text{zb}}_2t}  \right]\label{EQ:TrExp}
\end{align}
The second term in Eq.~\eqref{EQ:TrExp}, with the angular frequency $\omega_2^{\text{zb}}$, is easily identified as position-\ZB.
It is the result of an overlap between the components of positive and negative energies and of opposite spin/helicity clearly visible in the expectation value of $\hat{\alpha}_j$, $j=y,z$. 
The transverse position-\ZB~frequency $\omega^{zb}_2$ has the same properties, as the one discussed in the context of spin-\ZB~and, in particular, it is subjected to the blue-shift effect.

The first term in Eq.~\eqref{EQ:TrExp} with the angular frequency equal to the Larmor frequency $\omega_L$ is harder to be identified as traditional \ZB.
While it describes a phenomenon similar to traditional \ZB~--- oscillation in the expectation values of velocity and position, it does not stem from interference between solutions with positive and negative energy.
As it is clear from the structure of the matrix element $\langle\uparrow,l| \hat{\alpha}_j|l,\downarrow\rangle$ ($l=\pm$) of the velocity operator $\hat{\alpha}_j$ ($j=y,z$), it stems from matrix elements with the same sign of energy but with opposite spin orientations.
Explicitly, the matrix elements $\langle\uparrow,l| \hat{\alpha}_j|l,\downarrow\rangle$ are given by
\begin{subequations}
\begin{align}
\langle-,\downarrow|\alpha_y|-,\uparrow\rangle &+ \langle+,\uparrow|\alpha_y|+,\downarrow\rangle = \frac{cp_x(\hbar\omega_L-2\Delta)}{i\zeta} \nonumber \\
& + \frac{cp_x(\hbar\omega_L+2\Delta)}{i\eta},\\
\langle-,\downarrow|\alpha_z|-,\uparrow\rangle &+ \langle+,\uparrow|\alpha_z|+,\downarrow\rangle = \frac{cp_x(\hbar\omega_L-2\Delta)}{\zeta} \nonumber \\
& - \frac{cp_x(\hbar\omega_L+2\Delta)}{\eta},
\end{align}
\end{subequations}
where $\zeta = 2\sqrt{E_-^{\uparrow}E_-^{\downarrow}(E_-^{\uparrow}+E_0^{\uparrow})(E_-^{\downarrow}+E_0^{\downarrow})}$
 and $\eta =2\sqrt{E_+^{\uparrow}E_+^{\downarrow}(E_+^{\uparrow}+E_0^{\uparrow})(E_+^{\downarrow}+E_0^{\downarrow})} $.

Since $\zeta \gg \eta$ the amplitude of the oscillations will be dominated by the matrix elements $\langle-,\downarrow|\alpha_j|-,\uparrow\rangle$ ($j=y,z$) between the solutions with negative energy.
Furthermore, the matrix elements depend on the size of the relativistic Larmor frequency $\omega_L$, the interaction energy $\Delta$ and the momentum of the particle.
In the rest frame of the particle ($p_x=0$) the oscillations disappear.
They also disappear when the interaction energy $\Delta$ between the particle MDM/EDM and external static field is zero. 
This is a feature of relativistic spin splitting.
The lifting of spin degeneracy leads to precessionary motion in the plane transverse to the direction of spin splitting.
This shows that by lifting the spin degeneracy and by the orbital effect of spin splitting it might be possible to observe \ZB~with a frequency smaller than the limiting frequency ${2mc^2}/{\hbar}$ of a free particle.

The combination of two frequencies in the expectation values of $\langle\hat{r}_j\rangle$ leads to the appearance of a beating pattern with a beating frequency $\omega^{ob}_2=\omega^{zb}_2-\omega_L$.
It has the same properties as the beating frequency of the modified spin-\ZB.

The lowering of the \ZB~frequency $\omega^{\text{zb}}_2$ below the free particle limit ${2mc^2}/{\hbar}$, as well as the beating frequencies, lay down a route for eventual detection of \ZB~phenomena in conventional experiments.
This is further helped by the red-shift effect for the longitudinal beating frequency $\omega^{\text{ob}}$.
An alternative route for experimental studies of the above effects is quantum simulations.
In a previous work \cite{TenevA}, we have proposed the emulation of the Dirac equation with EDM in trapped ions.
The same scheme of simulation can be applied to the emulation of the effects considered here.
After the preparation of an appropriate initial state the dynamics can be simulated in the proposed way~\cite{TenevA}.
The different \ZB~frequencies can then be measured using the technique of mapping spatial variables onto the internal states and subsequent measurements of them \cite{Gerritsma}.

\section{Conclusions}

In this work we pointed out the existence of a \emph{motional blue-shift effect} for the \ZB~frequency of a free particle.
We predicted the existence of a ``forbidden frequency'' of spin precession in 1D and showed that it is not modified by the presence of external fields.
We also predicted a modification of the frequency of spin-\ZB.
We showed that the beating frequency stemming from the combination of spin-\ZB~and Larmor precession exhibits a \emph{motional blue-shift effect}.
Furthermore, we predicted the existence of two different position \ZB~phenomena: transverse and longitudinal.
The longitudinal position-\ZB~is characterized by two frequencies different from the free particle one ${2mc^2}/{\hbar}$ and leading to a beating pattern.
We showed that the two \ZB~frequencies exhibit a \emph{motional blue-shift effect}, while the beating frequency exhibits a \emph{motional red-shift effect}.
The transverse \ZB~is also characterized by two frequencies and a beating pattern similar to the spin-\ZB.
However, while the higher frequency can be interpreted as \ZB~frequency, the lower can not be identified as conventional \ZB, because it stems from matrix elements with the same sign of energy but opposite spin orientations.
We referred to it as orbital effect of spin splitting.

This work has been supported by the EU COST project IOTA and the Bulgarian NSF grants D002-90/08 and DMU03/107.

%

\end{document}